# Optogenetic control of cell signaling pathway through scattering skull using wavefront shaping

Jonghee Yoon[1,†], Minji Lee[2,†], KyeoReh Lee[1], Nury Kim[2], Jin Man Kim[3],
Jongchan Park[1], Hyeonseung Yu[1], Chulhee Choi[4], Won Do Heo[2,5,*] and YongKeun Park[1,*]

[1]*Department of Physics, Korea Advanced Institute of Science and Technology, Daejeon 305-701, Republic of Korea*

[2] *Department of Biological Sciences, Korea Advanced Institute of Science and Technology, Daejeon 305-701, Republic of Korea*

[3]*Graduate School of Medical Science and Engineering, Korea Advanced Institute of Science and Technology, Daejeon 305-701, Republic of Korea*

[4]*Department of Bio and Brain Engineering, Korea Advanced Institute of Science and Technology, Daejeon 305-701, Republic of Korea*

[5]*Center for Cognition and Sociality, Institute for Basic Science (IBS), Daejeon 305-811, Republic of Korea*

[†]*These authors contributed equally to this work.*

*Correspondence should be addressed to Y.-K.P (yk.park@kaist.ac.kr) or W.D.H (wondo@kaist.ac.kr)*

**We introduce a non-invasive approach for optogenetic regulation in biological cells through highly scattering skull tissue using wavefront shaping. The wavefront of the incident light was systematically controlled using a spatial light modulator in order to overcome multiple light-scattering in a mouse skull layer and to focus light on the target cells. We demonstrate that illumination with shaped waves enables spatiotemporal regulation of intracellular $Ca^{2+}$ level at the individual-cell level.**

Recently research has indicated significant potentials of optogenetics, light-controlled regulation of cell functions exploiting genetically modified light-sensitive proteins[1,2]. These have been demonstrated in various applications ranging from neuroscience[3] and cardiology[4], to genetics[5]. Recently, a significant breakthrough in optogenetics has been achieved by *in vivo* functional studies[3,4]. However, attempts toward *in vivo* optogenetics have been stymied by a fundamental limitation—light scattering. Multiple light scattering significantly limits light delivery through turbid media such as brain or skull layers (**Fig. 1a**). Consequently, existing *in vivo* optogenetic approaches rely on invasive methods including cranial windows involving skull removal or thinning[6], and invasive implementation of an optical fiber[7] (**Fig. 1b**). Although these methods extend the accessibility of optogenetics to deep tissue regions, unwanted cellular activities and tissue damage are inevitable. Recently developed red-shifted light-sensitive proteins, which absorb near-infrared (NIR) light[8], showed deep light penetration through biological tissues, but penetration depth of NIR light is still limited to a few millimeters due to multiple light scattering.

To deliver a controlled light through highly scattering tissues towards non-invasive *in vivo* optogenetic control, we exploit wavefront shaping techniques (**Fig. 1c**). Without controlling the incident wavefront, light diffuses out as it propagates through optically inhomogeneous complex media such as tissues, resulting in the formation of speckle patterns. This speckle pattern seems stochastic due to the complex distribution of intensity patterns. However, the speckle formation results from light interference, which is a deterministic process and thus can be systematically controlled by changing the incident wavefront[9]. We show that the shaped wavefront of an incident excitation light generates optimized focus through 300 μm-thick mouse skull. This enables the spatiotemporal regulation of intracellular $Ca^{2+}$ level of the irradiated cell, by activating the light-sensitive proteins.

**Results**

**Focusing through a highly scattering skull layer using wavefront shaping**

In order to demonstrate that the wavefront shaping of excitation beams enables the generation of selective focus through a highly scattering skull layer and thereby the optogenetic regulation of signaling-pathways in individual cells, we used an *in vitro* model system in which a spatial light modulator (SLM) was employed for wavefront shaping (**Fig. 2a**). The experimental setup consists of two parts: the wavefront shaping part for activating light-sensitive proteins and the fluorescence imaging part for measuring cellular responses (**Fig. 2b**). To demonstrate the capability of optogenetic control through scattering tissues, a skull layer dissected from a mouse without a thinning procedure was prepared. The dissected skull layer was then attached to the bottom of a culture plate in which cells were dispersed in a monolayer. Thickness of the dissected skull was approximately 300 μm, measured from the second harmonic generation image (**Fig. 2c**). This was thick enough to scramble the incident laser beam into speckle patterns.

To systematically and precisely generate focus though a turbid medium, we used an algorithm modified from parallel optimization[10]. The parallel algorithm optimizes half a field of SLM pixels using the other half as a reference. After optimizing one-half of the SLM pixels, light intensity at the focus is increased to the half-maximum intensity of the optimized focus. However, increased intensity at the target region after optimizing half of the SLM pixels could excite light-activating proteins and activate unwanted cellular responses. This would disturb the experimental results. Alternatively, we optimized both halves of the SLM pixels separately, and then the optimal SLM pattern was calculated by adding the relative phase value to one half of the optimized SLM pixels, while the values of the other half of the SLM pixels remained constant (**Supplementary Fig. 1**). This modified parallel algorithm prevents activation of cellular responses during the optimization process. With the optimized SLM pixels, optical focus of the excitation laser was formed clearly after passing through the brain skull layer, whereas an uncontrolled wavefront only generate diffused light or a speckle pattern (**Fig. 2d**). The FWHM size of the generated focus was 1.905 ± 0.224 μm, which is small enough to regulate optogenetic components as the subcellular resolution.

In principle, the intensity enhancement of an optimized focus $\eta$ is proportional to the number of SLM pixels used $N$ as, $\eta = (\pi/4) \cdot (N - 1) + 1$.[11] The use of large number of SLM pixels, and thus the increased optimization time, makes the optimization process vulnerable to the addition of more noise, which is obviously undesirable for *in vivo* applications. In order to reduce optimization time while maintaining the desirable degree of control for wavefront shaping, we grouped SLM pixels into $N$ equally sized square segments. For the experiments, we used $N = 55$ and 110, and the corresponding optimization times were 84.79 s and 140.05 s, respectively. The measured $\eta$ for experiments with the brain tissue only, were 52 ($N = 110$) and 28.4 ($N = 55$) (**Fig. 2e**). However, $\eta$ was decreased with live cells located on the top of the brain tissue: $\eta = 30.6$ ($N = 110$) and 14.5 ($N = 55$). Under live-cell conditions, cellular and intercellular movements, and fluctuation of the culture medium, may adversely affect the optimization process. Nonetheless, $\eta = 14.5$ is high enough for the spatiotemporal control of light-activating proteins *in vitro*.

To investigate the relation between skull thickness and the enhancement in intensity of optimized foci, we stacked mouse skull layers and then measured the intensity of optimized foci (**Fig. 2f**). We found that $\eta$ was relatively constant regardless of skull thickness: $\eta = 46.0$ (1 skull layer), $\eta = 46.5$ (2 skull layers), and $\eta = 42.5$ (3 skull layers). However, the measured total transmittance $T$ and intensity of optimized foci were decreased as skull thickness increases. Measured $T$ were 54.9%, 20.2%, and 10.5%, and intensity of optimized foci were 3342 ± 748, 359 ± 37, and 170 ± 23 (arbitrary unit), for 1, 2 and 3 skull layers, respectively. (**Fig. 2f**). Theoretially, $T$ in the diffusive regime ($L \gg l_s$) is can be written as follows[12,13],

$$T \cong \frac{l_s}{L}\left[1 + \frac{2(1+R)}{3(1-R)}\right], \qquad (1)$$

where $L$ is the thickness of a turbid medium, $l_s$ is mean free path, and $R$ is total reflectance. This relation indicates that $T$ is inversely proportional to thickness of a turbid medium. In our experiments, the measured $T$ and intensity of optimized foci through mouse skull were inversely proportional to skull thickness, which is consistent with theoretical expectation from Eq. (1) (**Fig. 2f**). Next, we measured the transport mean free path (TMFP) of frontal and parietal region of a mouse skull according to mouse strains and ages. The TMFP is defined as the average distance in which light completely losses its original propagation direction, which provides intuitive understanding of the degree of turbidity of complex media[14]. TMFPs of frontal and parietal bones were measured as 0.34 ± 0.13 and 0.15 ± 0.03, respectively (**Fig. 2g**). Measured TMFPs show significant difference between frontal and parietal bones, but no difference according to mouse strains and ages (**Supplementary Table 1**). Measured TMFPs are also consistent with previous reports [14,15].

**Spatiotemporal regulation of intracellular $Ca^{2+}$ activity with a shaped wave irradiation**

To test whether an optimized focus through the skull enables spatial control of light-sensitive proteins, we performed *in vitro* experiments using HeLa cells expressing optoFGFR1. OptoFGFR1 is a photoactivatable receptor tyrosine kinase which is activated by blue light (488 nm), and induces $Ca^{2+}$ release from the endoplasmic reticulum through downstream signaling pathway[16]. To measure the intracellular $Ca^{2+}$ level, R-GECO1, which emits red fluorescence by binding of $Ca^{2+}$ ions, was co-expressed with optoFGFR1 in HeLa cells. First, the mouse skull was irradiated using an uncontrolled plane wave (6 μW) through a high-NA objective lens, for 60–200 s. The beam transmitted through the mouse skull exhibited a speckle pattern, and all the cells inside the field of view were irradiated by the speckle light field (**Fig. 3a**). After illumination with the plane wave, not only the target cell but also adjacent cells showed overall increases in R-GECO1 signals (**Fig.**

**3b**). Then, we optimized the wavefront of the incident beam in order to focus the beam in the middle of the target cell, 15 min after the plane wave illumination (**Fig. 3a**). Illumination with the optimized shaped wave, with less power (2 μW), specifically induced R-GECO1 signal increases only in the target cell; adjacent cells did not exhibit significant changes in R-GECO1 signals (**Fig. 3c**). These results demonstrate that the wavefront shaping method enables spatial control of light-sensitive proteins at the subcellular resolution to specifically activate individual single cells through the highly scattering skull layer.

Temporal regulation and reversibility of optoFGFR1 activity were also accessed by repeated irradiations with the shaped wave onto HeLa cells behind the skull tissue. First irradiation of the target cell with the shaped wave (2 μW) increased R-GECO1 fluorescence signals only in the target cell (**Fig. 3d**). After withdrawal of light irradiation, we waited for an additional 18 min to recover the activity of optoFGFR1 in the irradiated cells. Then, we optimized a wavefront of the incident light again, and the R-GECO1 fluorescence signal in the same target cell was elevated again upon the second irradiation with the shaped wave (**Fig. 3e**). The levels of cellular response were similar in both irradiations with the shaped wave.

**Discussion**

In summary, this study provided the first *in vitro* demonstration of optogenetic control through a highly scattering skull layer using the wavefront shaping technique. By controlling the wavefront of an excitation beam impinging through the skull layer using a SLM, we experimentally demonstrated that optoFGFR1 expressing cells lying beyond the skull could be regulated spatiotemporally. This resulted in optical control of increased intracellular $Ca^{2+}$ levels in the activated cells.

In its present embodiment, the wavefront shaping was performed using transmission geometry. In the future, for *in vivo* applications involving live animals, the present method should be implemented using reflection geometry. This may require additional techniques such as optical phase conjugation[17,18], photo-acoustically tagged light[19], and wavefront-shaping optical coherence tomography[20]. Another limitation in the current study is the long optimization time due to the slow refresh rate of SLMs. Alternatively, dynamic mirror devices or deformable mirrors could be used to expedite the optimization process. Nonetheless, the present work provided the first demonstration of the importance of wavefront shaping techniques towards non-invasive *in vivo* optogenetic studies. Furthermore, the wavefront shaping approach provide large degrees of freedom in controlling the light, allowing the control of polarization[21], wavelength[22], spatiotemporal focusing[23], coherence gating for depth selection[20,24], and sub-wavelength information[25,26]. In addition, approaches based on scattering matrices[14,26-30] can also be applied to realize full-field optogentic control. Our approach allows an improvement in the applicability of optogenetics and should find widespread applications in *ex vivo* tissue models and for non-invasive regulation of light-sensitive proteins *in vivo*.

**Methods**

**Cell preparations**

HeLa cells were cultured in Dulbecco's modified Eagle's medium (DMEM) (catalog no. 11995-065, Gibco) supplemented with 10% fetal bovine serum (catalog no. 26140-079, Gibco) at 37°C with 10% $CO_2$. HeLa cells were transfected with optoFGFR1 and R-GECO1 by using Neon transfection system (catalog no. MPK5000, Invitrogen) according to manufacturer's instruction. For live cell imaging, transfected cells were plated on 40 mm dish (catalog no. 93040, TPP) and serum-starved with serum-free DMEM for 6hr. After starvation, all experiments were performed in a dark condition for protection HeLa cells from light. OptoFGFR1 construct was used as previously reported[16]. R-GECO1 plasmid was a gift from Robert Campbell (Addgene plasmid # 32444).[31]

**Mouse skull preparation**

For mimicking intact brain, we used a whole dissected mouse skull. All experiments used 7- to 10-week-old male Balb/c mice (Orient Bio Inc., Republic of Korea). The sample preparation procedures and the methods were approved by the Institutional Review Board (IRB project number: KA2013-15) and all the experiements were performed in accordance with the approved guidelines and regulations. The head was removed after euthanizing mouse. The scalp and muscles were eliminated using stainless-steel scissors. Then, the skull was gently dissected with a rongeur heading from the neck toward the nose. The dissected mouse skull was immersed in phosphate-buffered saline (Welgene Inc., Republic of Korea) for 24 hours to remove blood. Before experiments, the dissected mouse skull was dried and then attached at the bottom of the culture dish using a transparent double-sided tape. The transmittance and reflectance of the sample were measured using an integrating sphere. Then the reduced scattering coefficient was calculated from the inverse-adding doubling method.[32] The reduced scattering parameter of the skull was measured as 3.297 $mm^{-1}$.

**Optical setup**

Optical setup is composed of two parts; the wavefront shaping part for the excitation beam and the fluorescence imaging part for measuring R-GECO1 signals. A continuous-wave laser diode ($\lambda$ = 488 nm, 60 mW, Cobolt) is used to activate optoFGFR1 which absorbs a wavelength of 488 nm. A beam from the laser diode is expended by 4f lens system to cover whole active area of the SLM before passing through a half-wave plate (WPH10E-488, Thorlabs) and a polarizer (LPVISE100-A, Thorlabs). The SLM (X10468-01, Hamamatsu Photonics Inc. Japan) is placed at a conjugate image plane of the mouse skull. A CCD camera (Lt365R, Lumenera Inc., USA) is used as a detector for both the wavefront shaping and the fluorescence signals. For measuring fluorescence images, a LED ($\lambda$ = 565 nm, 880 mW, Thorlabs) is used to excite the R-GECO1 of which excitation and emission peaks are 561 nm and 600 nm, respectively. Bandpass filters are used as excitation (FF01-561/14-25, Semrock, Rochester, NY, USA) and emission (FF01-607/36-25, Semrock, Rochester, NY, USA) filters

**Wavefront shaping process**

In order to optimize a wavefront of the incident light, we used an algorithm modified from the parallel wavefront optimization method[10] to utilize the entire active area of the SLM for wavefront shaping. In our system, the SLM is directly projected onto the mouse skull; in other words, each SLM segments has a one-to-one correspondence to a positon on the mouse skull. The SLM pixels were grouped into $N$ ($N$ = 55, 110) equally sized square segments to avoid oversampling situation, and to reduce calculation time. Wavefront shaping procedure consists of three steps (Supplementary Fig. 1a). First, the one half (left panel) of the SLM segments is modulated, while the other half (right panel) of the SLM segments is fixed with a constant phase value. After finishing optimization of the left panel of the SLM segments, optimized phase values are recorded. Then, the phase values of the left panel of the SLM segments was reset to a constant phase (usually zero), while modulating the right panel of the SLM segments. After optimization of both panels of the SLM segments, the phase matching process was performed in order to find the value of a relative phase difference between two panels (Supplementary Fig. 1b). The optimized phase values are applied to both left and right panels. Then by monitoring the intensity of the generated focus, the left half of the SLM segments is fixed and the right panel of the SLM segments is added with a constant phase value between 0 and $2\pi$. The phase value which gives the maximum intensity is selected for finding the optimized SLM segments.

**Image analysis**

All image analysis were performed using Matlab and ImageJ softwares.

**ACKNOWLEGEMENTS**

This work was supported by from National Research Foundation (NRF) of Korea (2014K1A3A1A09063027, 2013M3C1A3063046, 2012-M3C1A1-048860, 2014M3C1A3052537), APCTP, KAIST-Khalifa University, KAIST EndRun-project.

**AUTHOR CONTRIBUTIONS**

Y.P, C.C, and W.D.H conceived the idea and directed the work. J.Y., K.L., and J.P. designed optics system and developed algorithms for experiments. M.L., N.K, and J.M.K established *in vitro* model. H.Y. measured optical parameters of skulls. All authors wrote the manuscript.

**COMPETING FINANCIAL INTERESTS**

The authors declare no competing financial interests.

**Figure legends**

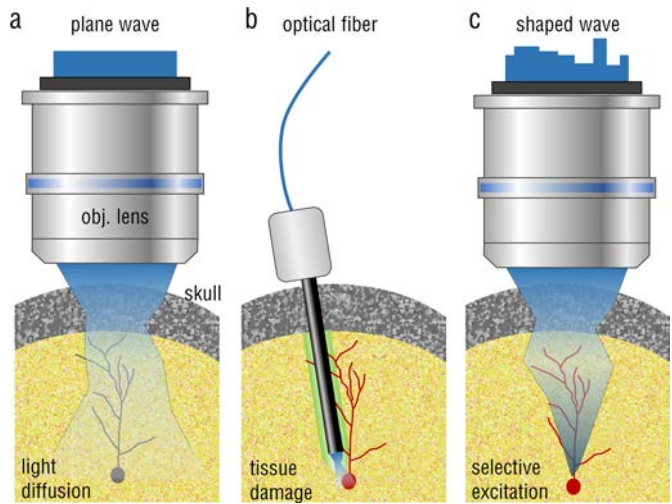

**Figure 1 |** Schematics of current optical delivery methods for *in vivo* optogenetics and the proposed concept: (a) Focusing with a conventional lens results in light diffusion due to multiple light scattering in a skull and brain tissues. (b) Irradiation of target regions using an implanted optical fiber is invasive and causes tissue damage (green color). (c) Illumination with a shaped beam can enable light focusing inside tissues without invasive processes such as skull thinning or optical fiber implantation.

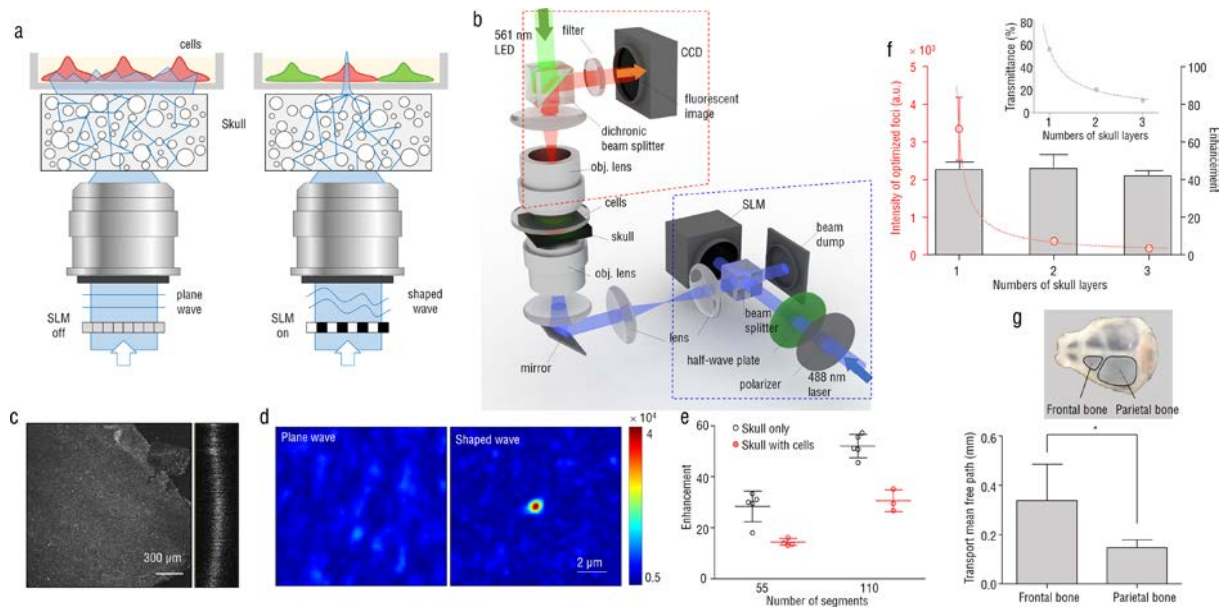

**Figure 2 |** Optical focusing through a skull layer using wavefront shaping technique: (a) Schematic of the experiment. Cells expressing both optoFGFR1 and R-GECO1 are placed above the skull layer. When illuminating with a plane wave through an objective lens, the beam undergoes multiple light scattering as it propagates through the skull and this diffused light may activate optigenetic signals in all cells in an

uncontrolled manner (left). By shaping the wavefront of the impinging beam using SLM, optical focus can be generated at the plane of the cells, thus activating optogenetic regulation at the level of individual cells (right). (b) Experimental setup, L1-3 lens: The blue dashed box indicates a wavefront shaping part, and the red dashed box indicates a fluorescence-imaging part. (c) Confocal image of second harmonic generation signals in the used mouse skull on the x-y plane (left) and in the axial direction (right). Scale bar 500 μm. (d) Intensity images of the transmitted beam through the skull without wavefront shaping (left), and with the wavefront shaping (right). (e) Measured intensity enhancements according to segment numbers and samples. (f) Measured intensity of optimized foci ($n = 5$) and intensity enhancements ($n = 5$) according to numbers of skull layers. The graph inset illustrates the optical transmittance according to numbers of skull layers. The circles represent experimental data, and a dashed line is a theoretically expected from Eq. (1). (g) Measured transport mean free path of the frontal ($n = 4$) and parietal bones ($n = 4$), respectively. Schematic of a mouse skull (top) and corresponding transport mean free paths of the mouse skull (bottom).

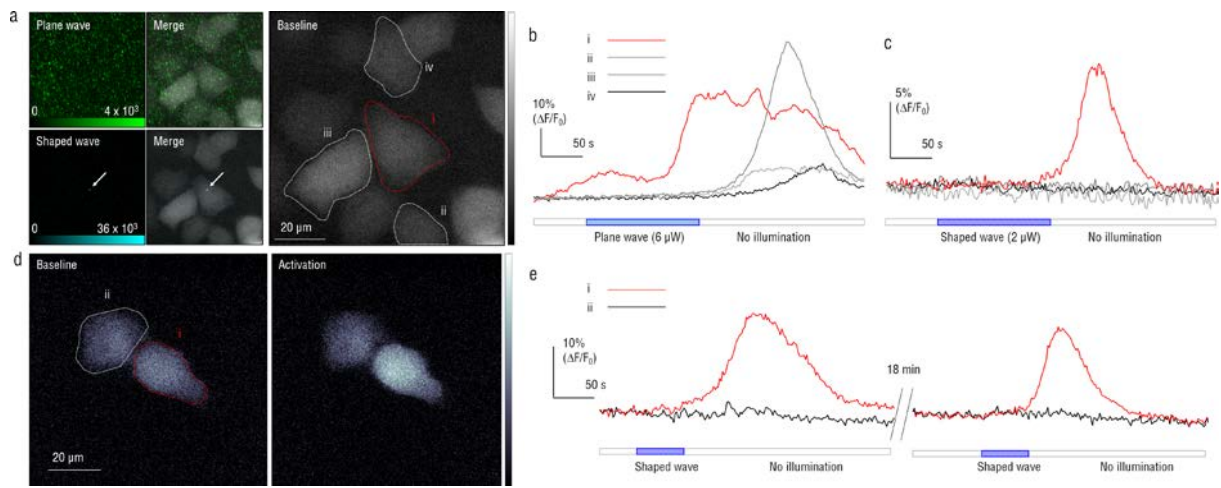

**Figure 3** | Spatiotemporal regulation of intracellular $Ca^{2+}$ level in HeLa cells co-expressing optoFGFR1 and R-GECO1, using wavefront shaping: (a) Targeted activation of $Ca^{2+}$ signaling in HeLa cells after focusing through the skull. Gray color indicates R-GECO1 fluorescence signals. Green indicates excitation beam intensity through the skull without wavefront shaping, and cyan indicates optimized focus with wavefront shaping. The white arrow indicates the location of the optimized focus. The dashed lines indicate boundaries of individual cells. The red dashed line indicates the target cell. Quantitative analysis of R-GECO1 fluorescence signals obtained from cells i–iv in (a), with a plane wave (b), and with the shaped wave (c). The blue bar and the blue checked bar indicate irradiation time of a plane wave and the shaped wave, respectively. (d) Reversible control of $Ca^{2+}$ level induced by optoFGFR1 in HeLa cells with repeated illumination using a shaped wave. Representative images show the baseline (left, no illumination) and maximal (right, shaped-wave illumination) R-GECO1 fluorescence signals. The dashed lines indicate each cell boundary. Red indicates the target cell. (e) Quantitative analysis of R-GECO1 fluorescence signals obtained from cells in (d). The cells were repeatedly illuminated using shaped waves. Each color matches the cells in (d). The blue-checked bars indicate irradiation time of the shaped wave.